# Coexistence of surface and bulk state and negative magnetoresistance in Sulfur doped $Bi_2Se_3$

Rahul Singh[1], Vinod K. Gangwar[1], D. D. Daga[1], Mahima Singh[1], A. K. Ghosh[2], Manoranjan Kumar[3], A. Lakhani[4], Rajeev Singh[1] and Sandip Chatterjee[1]

[1]Department of Physics, IIT(BHU), Varanasi-221005, India

[2]Department of Physics, Banaras Hindu University, Varanasi-221005, India

[3]S. N. Bose National Centre for Basic Sciences, Kolkata- 700098, India

[4]UGC-DAE Consortium for Scientific Research, Indore, Madhya Pradesh 452017, India

**Abstract**

The magneto-transport properties in Sulfur doped $Bi_2Se_3$ are investigated. The magnetoresistance (MR) decreases with increase of S content and finally for 7% (i.e. y=0.21) S doping the magnetoresistance becomes negative. This negative MR is unusual as it is observed when magnetic field is applied with the perpendicular direction to the plane of the sample. The magneto-transport behavior shows the shubnikov-de hass (SdH) oscillation indicating the coexistence of both surface and bulk states. The negative MR has been attributed to the bulk conduction.

**Introduction**:

Among the various discovered Topological Insulators (TIs) materials $Bi_2Se_3$ is one of the most promising candidates as it has a single Dirac cone in the Brillouin zone and relatively large bulk energy gap of 0.3eV, sufficient for room temperature applications [1,2]. Topological insulators will also be of interest for spintronic materials as the Dirac states can be used to carry the spin current with small heat dissipation [3,4].

Moreover, even though $Bi_2Se_3$ is arguably the most simple representative of the 3D TI family, accessing the topological surface states (TSS) in transport has been hindered by a large residual carrier density in the bulk [5,6]. While Shubnikov–de Haas (SdH) oscillations are a powerful means to distinguish between bulk and surface charge carriers, their analysis and interpretation remains controversial. The literature emphasizes the difficulty in distinguishing between the bulk, TSS, and a two dimensional charge-accumulation layer [5–11]. Apart from the TSS, the electronic bulk states in $Bi_2Se_3$ are of particular interest since their spin splitting is found to be twice the cyclotron energy observed in quantum oscillation [12,13] and optical [14] experiments. Another peculiar property of $Bi_2Se_3$ and other 3D TIs is the observation of a linear positive magnetoresistance (MR) that persists up to room temperature [15–20]. Under a perpendicular magnetic field, positive magnetoresistance (MR) effects have been widely observed in 3D TI systems, such as the weak anti-localization (WAL) effect [9–11] and the linear MR effect [12,13]. At low temperatures and high magnetic field, Shubnikov–de Haas (SdH) oscillations usually superpose on the positive MR background and the π Berry phase can be extracted, which shows the transport properties of 2D Dirac fermions [14–18]. Tang et al. [21] remarkably reveal that the linear MR is induced by a two-dimensional transport. He et al. [22] conduct a magnetotransport study on ultrathin film of $Bi_2Se_3$ with variable thickness grown by molecular beam epitaxy, which shows much smaller magnetoresistance (less than 10%) at 14 T. Yan et al. [23] have reported a large linear MR nearly 400% at low temperature and a corresponding high mobility of 10 000 $cm^2V^{-1}$ $s^{-1}$ in $Bi_2Se_3$ nanoplate synthesized via chemical vapor deposition (CVD) method. The linear MR persists even at room temperature with the value of 75%.

The recent interest on the unique topological properties in 3D massless Dirac fermions in “3D Dirac” or “Weyl” semimetals [24] can be revealed in magnetotransport experiments. Examples include the observation of an extremely large positive MR [25], linear MR [26], and, more specifically, the negative longitudinal MR (NLMR) predicted to appear in Weyl semimetals when the magnetic and electrics field are coaligned. In a recent theoretical study, however, it was proposed that the NLMR phenomenon may in fact be a generic property of metals and semiconductors [27], rather than something unique to topological semimetals. In the present investigation we have seen that with doping of S in $Bi_2Se_3$ the magnetoresistance (MR) gradually decreases and finally interestingly it shows NMR.

**Results & Discussions**:

Fig.1 shows the variation of Hall resistivity as a function of applied magnetic fields at different temperatures for $Bi_2Se_{3-y}S_y$ (with y= 0, 0.06, 0.15, 0.21). Slope of the curve is negative showing that carriers in pure and S doped $Bi_2Se_3$ are *n* type for the entire range of temperature of measurement. We have determined the mobility (μ) of the carriers from the Hall data. Calculated mobility as function of temperature is shown in the inset of Fig.1. It is observed that as we increase both the temperature and field, mobility decreases. This is due to the fact that with decreasing temperature, freezing out of phonons takes place and thus thermal vibration or the contribution of phonon decreases and high mobility prevails. Similar trend happens with the magnetic field. We have also estimated the carrier concentration from Hall data at low field. It is observed that carrier density for doped and undoped samples increases with temperature. Since, topological insulators are insulating in bulk but conducting on surface, at high temperature bulk contribution dominates over surface contribution which in effect enhances the carrier concentration. In fact, appearence topological surface state is a complete quantum phenomenon and therefore, existence of quantum mechanical behavior is significant at very low temperature. In consequence, at very low temperature surface state dominates over bulk state and that is why the carrier concentration is low at low temperature ($T \leq 20K$) and very large at high temperature. Moreover, the rate of increment of carrier density is also increasing with the increase of temperature; this also confirms that bulk insulating character is dominating over surface metallic character of the sample at higher temperature. The carrier concentration and mobility estimated from Hall data of $Bi_2Se_{3-y}S_y$ (with y= 0, 0.06, 0.15, 0.21) are shown in Table 1. It is observed that with S doping the mobility decreases and carrier concentration increases.

Fig.2 shows the longitudinal resistance vs temperature of $Bi_2Se_{3-y}S_y$ (with y= 0, 0.06, 0.15, 0.21) samples. The graph shows positive slope indicating their metallic behavior, as with increase of the S concentration, the resistivity decreases which might be due to the enhanced carrier concentration, caused by defects created at Se site as each Se vacancy donates two inherent electrons which can be described as:

$$Se_{Se} \rightarrow V_{Se^{\bullet\bullet}} + Se(g) + 2e'$$

The magneto-resistance (MR) as a function of a magnetic field at different temperature of $Bi_2Se_{3-y}S_y$ (with y= 0, 0.06, 0.15, 0.21) samples are shown in fig. 2. We have applied the magnetic field along the perpendicular direction of the plane of the sample. We have defined MR as [p(H)-p(0)]/p(0)*100%. In figure 2, we see that resistivity value increases with increase in magnetic field for y=0, 0.06 and 0.15 samples. $Bi_2Se_3$ shows a large linear MR nearly 200% at low temperature but when we increase the concentration of S its value decreases down to 7% as shown in figure 2(d), as increasing the carrier concentration lead to decrease in MR. Moreover, a negative magnetoresistance is observed for y=0.21 sample. For a high magnetic field, Landau-level induced SdH oscillations were observed at low temperatures. Quantum oscillations are clearly visible in the second derivative $-d^2\rho_{xx}/dB^2$, as a function of the inverse field as shown in fig.3. We have already mentioned that we have measured the MR at perpendicular magnetic field configuration only. Several frequencies are found in this perpendicular field measurement for all the samples. In order to identify the origin of the quantum oscillations, we have performed fast Fourier transforms (FFTs). It is clear from the fig.3 that only three frequencies are observed for all the samples . Among three frequencies one is for bulk and two are for the surface states. Taking the Onsager relation, i.e., the extremal cross section of the Fermi surface $A(E_F)=4\pi^2e/hF$ (F is the frequency) and assuming a spherical pocket we have calculated the $n_{bulk}$ for the bulk band corresponding to the pocket with the lowest frequency for all the samples. For the surface states, we have also estimated the carrier densities. All the bulk and surface carrier densities are shown in Table 2. From the quantum oscillation analysis, we have also calculated the total carrier concentration of $n_{tot,}^{SdH}$ (given in Table 2) which are in excellent agreement with $n_{Hall}$ and also consistent with those already reported [28].

The slope obtained from Landau-level fan diagram (Landau index vs. 1/B, B being the magnetic field) of each sample (shown in Fig. 3) reflects a 2D electron density of n= (e/h)BF. Additionally, the Landau level fan diagram shows an intercept at ~0.5 for the undoped sample, indicating that the Dirac fermions dominate the transport properties due to the additional Berry phase π. It is found that as the S content increases the deviation of the intercept from 0.5 also increases revealing that in the transport properties the contribution of Dirac fermion decreases, while the contribution of normal fermion increases. This clearly indicates that bulk conduction gradually dominates over surface conduction with S doping.

The observed negative magnetoresistance (NMR) in the y=0.21 sample is unusual. But the transition from positive to negative magnetoresistance is systematic. Generally, the NMR in TIs is observed when applied magnetic field is parallel to the electric current [28-29]. In the present investigation the NMR is found when the magnetic field is perpendicular to the electric current. The three possible reasons, *viz.*, Kondo effect quenching [30-31], transition from paramagnetic insulating to the ferromagnetic metallic state [32], and chiral anamoly as is observed in Weyl semimetal [33-35] are not the origin for the observation of NMR in the present case. The last one is expected to be observed only in the longitudinal configuration that is when magnetic field is parallel to the electric current. However, in the present investigation the magnetic field is perpendicular to the electric current. Moreover, NMR was also found in other Anderson localized electron systems explained by several different mechanisms [36] which is inconsistent with the metallic regime here.

So far, most of the reported NMR effects found in 3D TIs without magnetic doping are due to the weak localization (WL) effect coexisting with the weak anti localization effect under a low magnetic field [37]. However, our observed NMR cannot come from the WL effect for the following reasons. The NMR in the present investigation shows a weak temperature dependence in a wide range from 2 K to 20 K, which is not consistent with the weak localization from quantum interference because the phase coherence length should be sensitive to temperature. Furthermore, the WL induced NMR will saturate on increasing the magnetic field to~1 T as the magnetic length is smaller than the phase coherence length in these topological insulators [38,39]. However, our observed MR is still not saturated when the magnetic field is more than 3 T. Also, the NMR persists until 200K, far beyond the point at which a weak localization effect can exist. Furthermore, in a recent paper [40] it has been proposed that the observed NMR might be due to the Zeeman splitting where it has been pointed out that due to the helical spin and orbital angular momentum of the surface states, the additional Zeeman energy reaches maxima in the parallel direction with respect to the applied magnetic field. The rotational symmetry breaking of the Fermi circle results in spin polarization, that is, the in equal density of spin-up ($D\uparrow$) and spin-down ($D\downarrow$) surface electrons. But this mechanism is also not applicable here as we have applied the magnetic field in the perpendicular direction and also S is non-magnetic. However, we have carried out the DFT calculation (Fig.4) and we see that undoped sample shows some asymmetry in spin up and spin down state. When S is doped no asymmetry is observed and in consequence no local magnetic moment with S doping exists.

In a recent paper [40] Breunig et al. have reported the NMR under perpendicular magnetic field in $Bi_{2-x}Sb_xTe_{3-y}Se_y$ topological insulator and they have proposed that the observed NMR under perpendicular magnetic field is due to the electron puddles. But for electron or hole puddles to occur there should be thermally activated conduction. But in our case all the samples (undoped and S doped) show completely metallic behavior. This indicates that the surface state dominates over the bulk state even with the 7% S doping. Therefore, most likely, electron or hole puddles are not the origin of negative MR at perpendicular (to the current) magnetic field in y=0.7 sample.

It is clear from the Hall effect data that the Fermi level is located at the bulk conduction band due to the inevitable *n*-type doping from Se vacancies [41] which means that the bulk conduction electrons and the surface states can coexist to contribute to the conductance. Therefore, the bulk origin may play a dominant role in the NMR as it is observed under an perpendicular magnetic field, which is consistent with the three dimensional bulk conduction channels. To further support this we have fitted the MR data with the HLN formula [42]. According to the HLN formula, magneto-conductance can be expressed as

$$\sigma(B)= -A\ [\Psi(1/2 + \hbar /4e\ l_\varphi^2\ B)-ln(\hbar /4e\ l_\varphi^2)]$$

Where ψ is the digamma function, $l\varphi$ is the phase coherence length, the distance travelled by an electron with a constant phase. $A$ is related to the number of conduction channel in a sample, given by, $A=\alpha(e^2/2\pi^2\hbar)$ with α=1/2 per conduction channel. We have fitted our experimental data in the low field lange (0 to 1T) with the above equation for $Bi_2Se_{3-y}S_y$ (with y= 0, 0.06, 0.15, 0.21) (shown in Figure 5) and the fitted parameters $A$ and $l_\varphi$ are determined. Total numbers of channels have also been determined from the value of $A$. Temperature dependence of $l_\varphi$ can be expressed as [43, 44]:

$$1/\ l_\varphi^2(T)= 1/\ l_\varphi^2(0)+A_{ee}T +A_{ep}T^2$$

Here $l_\varphi$ is the zero temperature phase coherence length, $A_{ee}$ and $A_{ep}$ are respectively, electron-electron and electron-phonon interaction term. We have fitted the temperature dependent $l_\varphi$ with the above equation. The obtained fitting parameters are presented in Table 3. In figure 5 we have represented the number of channels and $l_\varphi$ as function of temperature. The numbers of channels are five times more than those found in two-dimensional systems and the values are consistent with those reported for single crystals [45]. The change of numbers of channels with S doping is also consistent with the resistivity behavior. It is observed in y=0.06 sample the number of channels decrease but as the doping concentration increases to y=0.15 the numbers of channels increase but still lower than those of the undoped sample. With further increase of doping concentration to 7% interestingly surface channel density decreases. We have observed in low temperature resistivity data (Fig.2) that initially conductivity decreases for y=0.06 sample and for y=0.15 the conductivity becomes higher than that of the y=0.06 sample but remains low compared to undoped sample. Furthermore, y=0.21, maximum conductivity is observed which is larger than that of the undoped sample. The enhancement of conductivity and decrement of surface channel density can be explained by the increment of number of bulk channels. Therefore, it confirms that the NMR is due to the bulk conduction.

**Conclusion**:

We have investigated the magneto-transport properties of $Bi_2Se_{3-y}S_y$ (with y= 0, 0.06, 0.15, 0.21). All the samples show the metallic behavior throughout the whole range of temperature measurement. Initially at lower temperature (below 100 K) the resistivity is greater than the undoped sample for y=0.06 and 0.15 samples. But as S content increases the resistivity decreases and finally for y=0.21 sample the resistivity becomes lowest throughout the whole temperature range of measurement. The MR also decreases with increase of S content and finally for y=0.21 sample it becomes negative. All the samples show SdH oscillations in the $d^2\rho_{xx}/dB^2$ as function of inverse magnetic field curves. The Fast Fourier transform of SdH oscillation shows the existence of both surface and bulk states. The NMR of y=0.21 sample has been explained as the dominance of the bulk conduction over surface conduction.

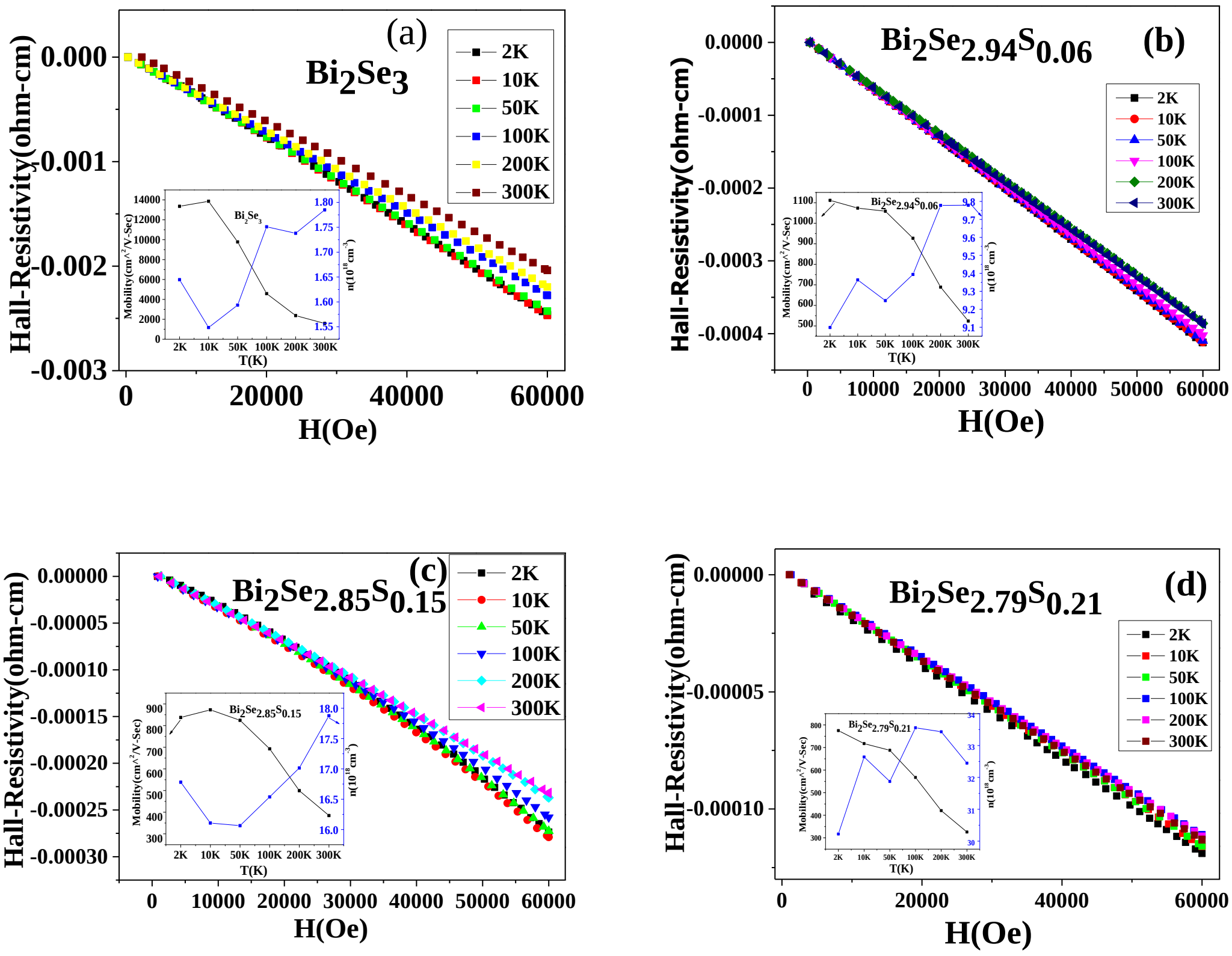

**Figure 1:** Hall resistivity as a function of magnetic field for $Bi_2Se_{3-y}S_y$ (with y= 0, 0.06, 0.15, 0.21) samples. Insets: variations of carrier concentration and Hall mobility as function of temperature.

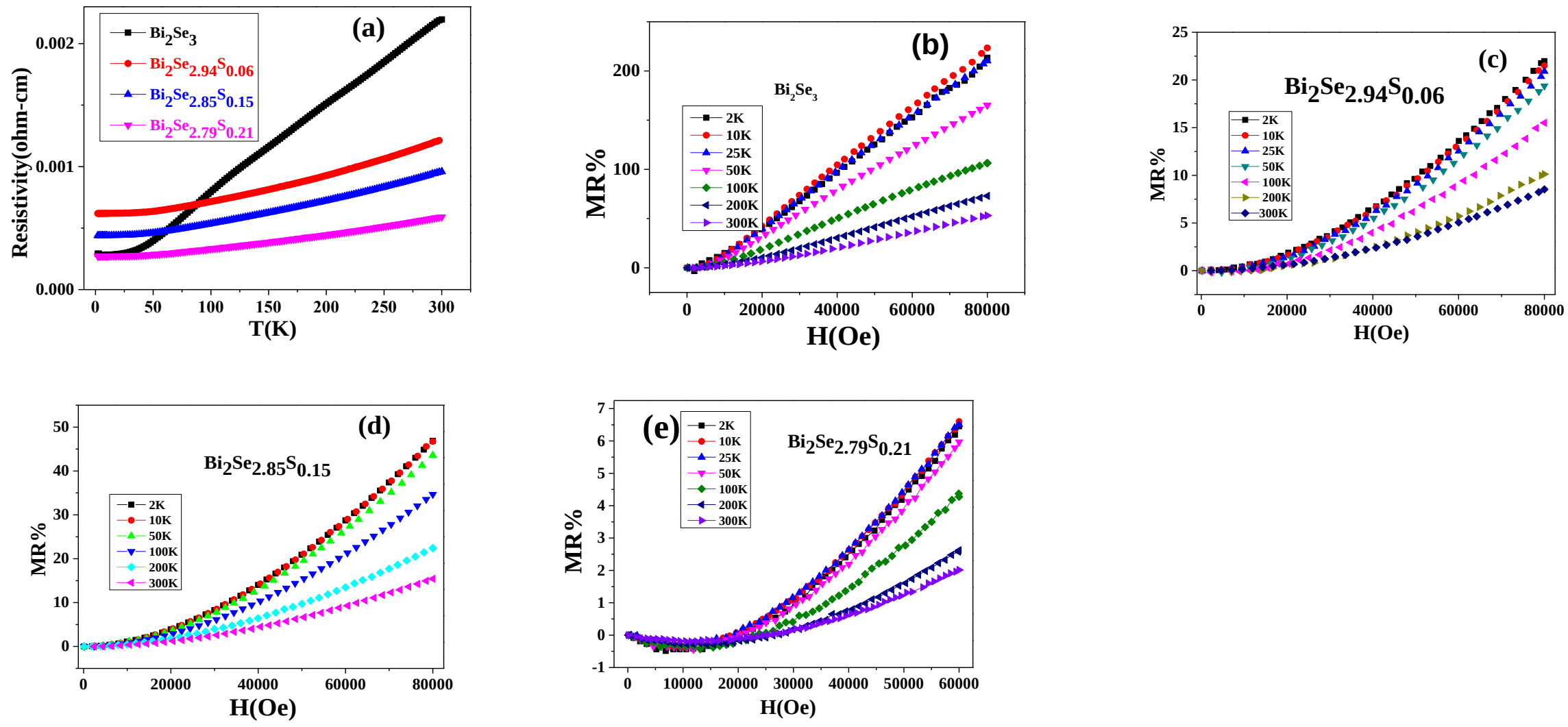

**Figure 2.** (a) Resistivity as function of temperature of $Bi_2Se_{3-y}S_y$ (with y= 0, 0.06, 0.15, 0.21) samples. (b-d) MR as function of magnetic field at different temperatures.

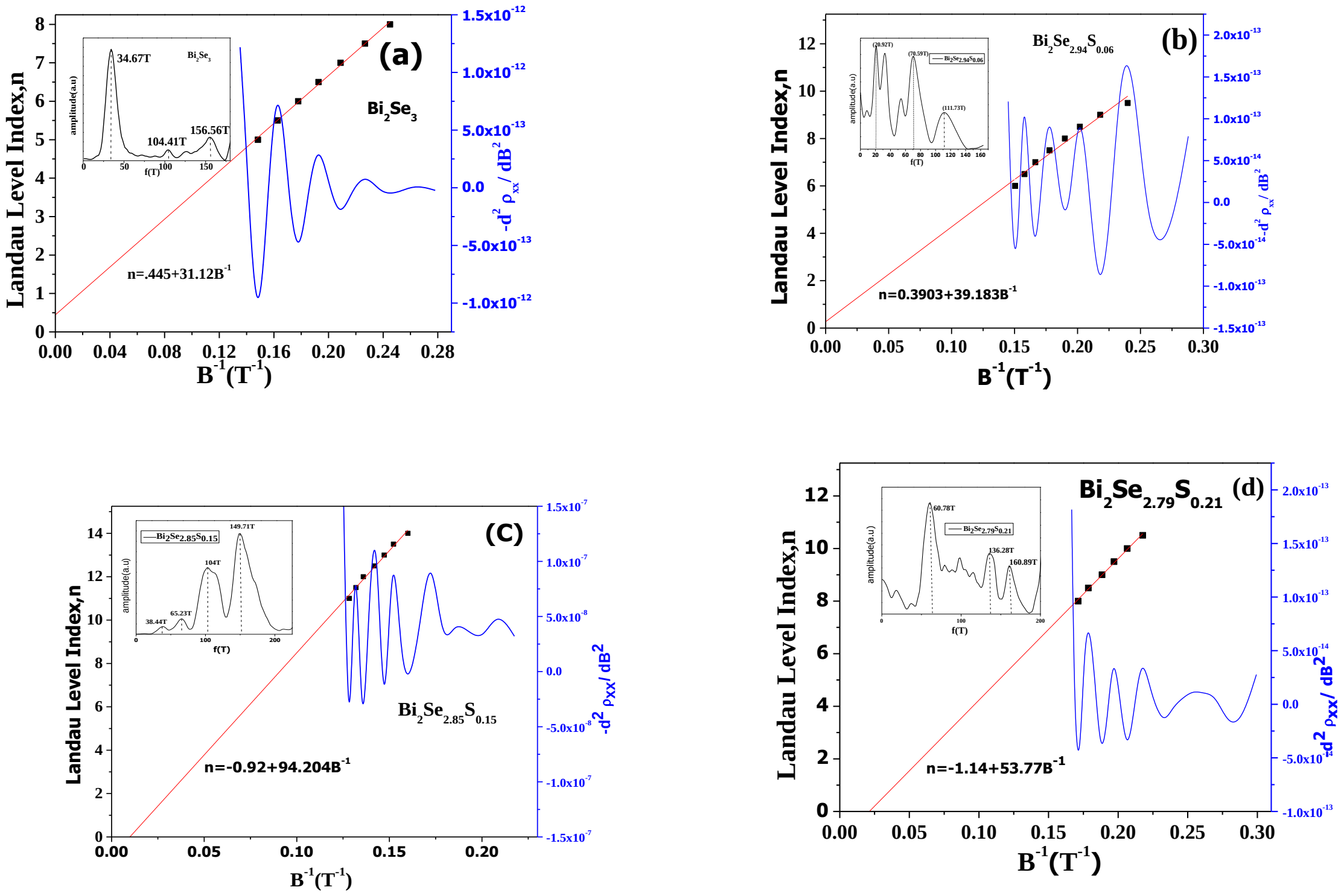

**Figure 3.** SdH oscillation of $Bi_2Se_{3-y}S_y$ (with y= 0, 0.06, 0.15, 0.21) samples shown from $d^2\rho_{xx}/dB^2$ as function of inverse magnetic field and Landau level index as a function of inverse magnetic field. Insets: Fast Fourier transform of the SdH oscillations.

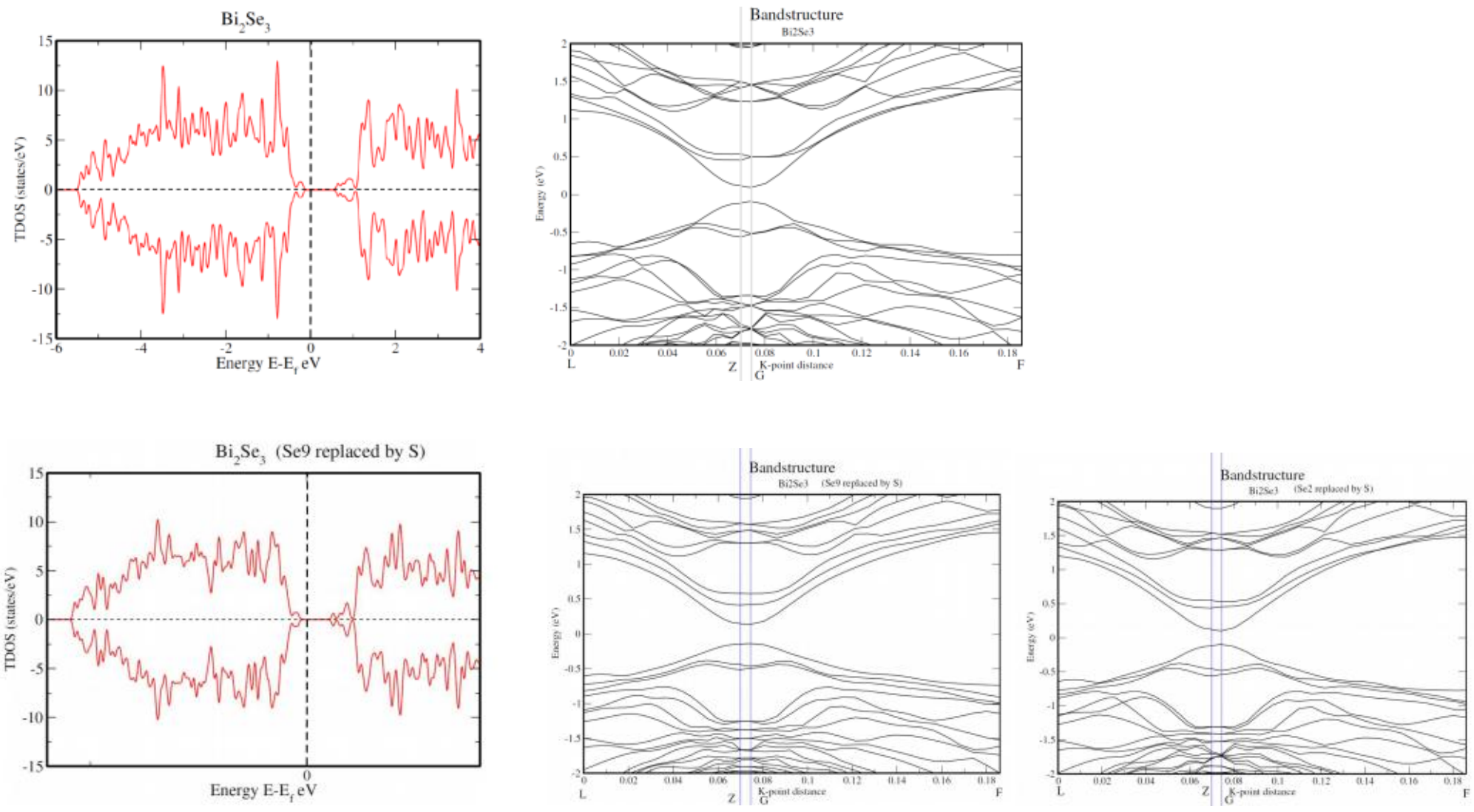

**Figure 4.** Total DOS of $Bi_2Se_3$ and S doped $Bi_2Se_3$. The dotted vertical line marks the Fermi energy. Band structure of of $Bi_2Se_3$ and S doped $Bi_2Se_3$ along k-points L-Z,Z-G,G-F

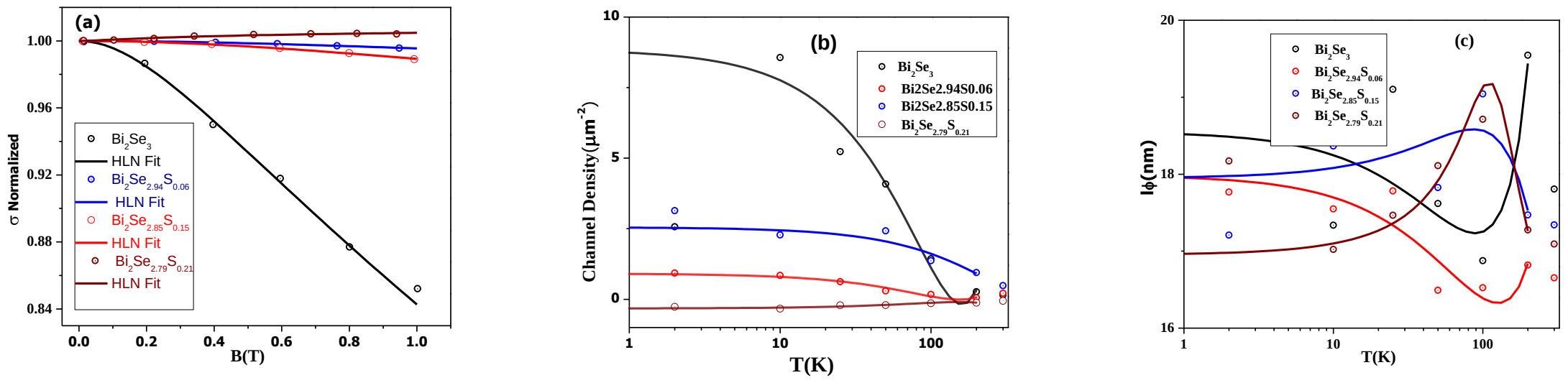

**Figure 5**. (a) Fitting of conductivity as function of magnetic field of $Bi_2Se_{3-y}S_y$ (with y= 0, 0.06, 0.15, 0.21) with HLN model. (b) Channel density of $Bi_2Se_{3-y}S_y$ (with y= 0, 0.06, 0.15, 0.21)as a function of temperature estimated from HLN model fitting. (c) Variation of coherence length of $Bi_2Se_{3-y}S_y$ (with y= 0, 0.06, 0.15, 0.21) as function of temperature estimated from HLN model fitting.

**Table-1:** Carrier concentration and Hall mobility $Bi_2Se_{3-y}S_y$ (with y= 0, 0.06, 0.15, 0.21) estimated from Hall data

| **Temperature** | **$Bi_2Se_3$ (n) $cm^{-3}$** | **$Bi_2Se_{2.94}S_{0.06}$ (n) $cm^{-3}$** | **$Bi_2Se_{2.85}S_{0.15}$ (n) $cm^{-3}$** | **$Bi_2Se_{2.79}S_{0.21}$ (n) $cm^{-3}$** |
|---|---|---|---|---|
| **2K** | 1.6447e+18 | 9.09847e+18 | 1.678e+19 | 3.02246e+19 |
| **10K** | 1.54848e+18 | 9.363414498e+18 | 1.6107e+19 | 3.264e+19 |
| **50K** | 1.5936e+18 | 9.247956e+18 | 1.6064e+19 | 3.187314e+19 |
| **100K** | 1.7510e+18 | 9.3939194e+18 | 1.6538e+19 | 3.355754e+19 |
| **200K** | 1.738e+18 | 9.777896e+18 | 1.7016e+19 | 3.3432158e+19 |
| **300K** | 1.785e+18 | 9.77914e+18 | 1.787754e+19 | 3.24482847e+19 |

| **Temperature** | **$Bi_2Se_3$ (μ) $cm^2$/V-Sec** | **$Bi_2Se_{2.94}S_{0.06}$ (μ) $cm^2$/V-Sec** | **$Bi_2Se_{2.85}S_{0.15}$ (μ) $cm^2$/V-Sec** | **$Bi_2Se_{2.79}S_{0.21}$ (μ) $cm^2$/V-Sec** |
|---|---|---|---|---|
| **2K** | 13357.234 | 1111.5744 | 837.67 | 774.958 |
| **10K** | 13862.2 | 1073.79093 | 872.84 | 717.419 |
| **50K** | 9779.537 | 1058.7183 | 824.417 | 687.76 |
| **100K** | 4583.51 | 927.0268 | 692.585 | 567.577 |
| **200K** | 2374.42 | 689.0339 | 499.533 | 420.554 |
| **300K** | 1587.46 | 523.6764 | 384.529 | 325.751 |

**Table-2**: Surface and bulk Carrier concentration of $Bi_2Se_{3-y}S_y$ (with y= 0, 0.06, 0.15, 0.21) estimated from SdH oscillation

| | **$n_b(cm^{-3})$** | **$n_s(cm^{-2})$** | **$n_s(cm^{-2})$** | **$n_s(cm^{-2})$** | **$n_{total}(cm^{-3})$** |
|---|---|---|---|---|---|
| **$Bi_2Se_3$** | 1.1514 e+18 @34.67T | 2.5197e+18@104.41T | 3.77782 e+17 @156.56T | | 1.15181 e+18 |
| **$Bi_2Se_{2.94}S_{0.06}$** | 5.3969e+17 @20.92T | 1.7016 e+12 @70.59T | 2.6964 e+12 @111.73T | | 5.399 e+17 |
| **$Bi_2Se_{2.85}S_{0.15}$** | 1.3442e+18 @38.44T | 1.5742 e+12 @65.23T | 2.5098 e+12 @104T | 3.6129 e+12 149.71T | 1.3446 e+18 |
| **$Bi_2Se_{2.79}S_{0.21}$** | 2.672665e+18 @60.78T | 3.2888 e+12 @136.28T | 3.8827 e+12 @160.89T | | 2.6731 e+18 |

**Table 3:** Different fitting parameters (Coherence length, electron-electron and electron-phonon interaction terms) obtained from HLN model fitting

| **Sample** | **lφ** | **Aee** | **Aep** |
|---|---|---|---|
| **$Bi_2Se_3$** | 1.95548322e+01 | 1.63087430e-05 | -8.37975321e-08 |
| **$Bi_2Se_{2.94}S_{0.06}$** | 1.72248007e+01 | 5.69569118e-06 | -1.74487415e-08 |
| **$Bi_2Se_{2.85}S_{0.15}$** | 1.80614107e+01 | -7.18906785e-06 | 5.05149193e-08 |
| **$Bi_2Se_{2.79}S_{0.21}$** | 4.91138372e+01 | -5.92905790e-06 | 2.71128996e-08 |